\documentclass[aps,prd,superscriptaddress,nofootinbib,eqsecnum]{revtex4}

\pdfoutput=1

\usepackage{amsfonts}
\usepackage{amsmath}
\usepackage{amssymb}
\usepackage{graphicx,color}
\usepackage{float}
\usepackage{hyperref}
\usepackage{subfigure}

\begin{document}

\title {Fluctuations through a Vibrating Bounce}

\author{Robert Brandenberger}
\affiliation{Physics Department, McGill University, Montreal, QC, H3A 2T8,
  Canada}

\author{Qiuyue Liang}
\affiliation{CAS Key Laboratory for Research in Galaxies and Cosmology,
  Department of Astronomy,
  University of Science and Technology of China, Chinese Academy of Sciences,
  Hefei, Anhui 230026, China}

\author{Rudnei O. Ramos} 
\affiliation{Departamento de F\'{\i}sica Te\'orica, Universidade do
Estado do Rio de Janeiro, 20550-013 Rio de Janeiro, RJ, Brazil}

\author{Siyi Zhou}
\affiliation{Department of Physics and Jockey Club Institute for Advanced
  Study, The Hong Kong University of Science and Technology,
  Clear Water Bay, Kowloon, Hong Kong, P.R.China}

\begin{abstract}

We study the evolution of cosmological perturbations in a non-singular
bouncing cosmology with a bounce phase which has superimposed
oscillations of the scale factor. We identify length scales for which
the final spectrum of fluctuations obtains imprints of the non-trivial
bounce dynamics.  These imprints in the spectrum are manifested in the
form of damped oscillation features at scales smaller than a
characteristic value and an increased reddening of the spectrum at all
the scales as the number of small bounces increases. 

\end{abstract}

 \maketitle

\section{Introduction}

Some bouncing cosmologies provide an alternative to cosmological
inflation as a way to obtain primordial cosmological fluctuations
(see, e.g., Ref.~\cite{BP} for a recent review). Specifically, in a
model which contains a matter-dominated phase of contraction, initial
vacuum fluctuations in the far past which exit the Hubble radius
during the matter-dominated contracting phase evolve into a
scale-invariant spectrum of cosmological perturbations~\cite{FB,
  Wands}. {}For example, in the case of an Ekpyrotic  contracting
universe~\cite{Ekp} entropy fields can source scale-invariant
curvature fluctuations~\cite{newEkp}. In all bouncing cosmologies, new
physics is required in order to obtain a non-singular cosmological
bounce. Such new physics could come from the matter sector (see, e.g.,
Refs.~\cite{matterBounce-1, matterBounce-2}), from modifications of
the classical gravitational  action (as for example in Horava-Lifshitz
gravity~\cite{HLbounce} or in the non-local gravity construction of
Ref.~\cite{Biswas}), or from quantum gravity effects. Examples of the
latter are the bounce in loop  quantum cosmology (see, e.g.,
Ref.~\cite{LQCbounce} for reviews), in deformed AdS/CFT
cosmology~\cite{Elisa},  the S-brane bounce of Ref.~\cite{Costas} and
the temperature bounce in String Gas Cosmology~\cite{BV}. 

Concerning the robustness of the computations of the spectrum of
cosmological fluctuations, an advantage of bouncing cosmologies
(without an inflationary phase after the bounce) is that the physical
length of modes which are probed in current observations remain in the
far infrared throughout the cosmological evolution as long as the
energy density at the bounce point is smaller than the Planck
density. Hence, the computations can be done in the realm where
effective field theory is well justified. This is in contrast to the
situation in inflationary cosmology~\cite{MB} where the physical
wavelengths of even the largest scales which are currently observed
are smaller than the Planck length at the beginning of inflation
(provided that the inflationary phase lasts slightly longer than the
minimal period it has to last in order to solve the horizon and
flatness problems of Standard Big Bang Cosmology). 

A key question is to whether the predictions for cosmological
perturbations at late times in the expanding phase are sensitive to
the details of the bounce phase. {}For simple parametrizations of the
bounce phase, detailed studies have shown that the spectral shape
does not change during the bounce phase provided that the duration of
the bounce phase is shorter than the length scale of the fluctuations
at the bounce point (see, e.g., Ref.~\cite{matterBounce-2} in the case
of matter-driven bounces, Ref.~\cite{HLflucts} in the case of the
Horava-Lifshitz bounce, Ref.~\cite{Elisa} in the case of the AdS/CFT
bounce, and Ref.~\cite{Subodh} for the S-brane bounce). On the other
hand, there are examples where the bounce phase yields dramatic
changes in the spectrum~\cite{BXue}. The reason why such dramatic
changes are possible is that the Hubble radius at the bounce point is
infinite, and we cannot invoke the freezing of cosmological
perturbations on super-Hubble scales to argue for a constancy of the
spectrum\footnote{Note that there are models in which the spectrum of
  scalar fluctuations is boosted {\textemdash} by a factor independent
  of wavelength on large scales {\textemdash} during the bounce
  phase~\cite{Jerome}.}.

To further analyze the sensitivity of the spectrum of cosmological
fluctuations on the details of the bounce phase, we here consider a
toy model where the scale factor undergoes small amplitude
oscillations during the bounce phase. Such a behavior may emerge from
certain models motivated by ideas from loop quantum
gravity~\cite{Alesci}.   Heuristically, one would argue that those
small bounces will not influence the large scale modes  provided that
the wavelengths of these modes are so large scale that they would not
feel the small  scale fluctuations of the scale factor. On the other
hand, smaller scale modes whose wavelength is comparable or smaller to
the total duration of the bounce phase should be sensitive to the
details of the dynamics during the bounce.  In this work, we would
like to give a careful treatment to see  if this is really the case. 

In the case of a cyclic cosmology, when the time interval between
cycles is larger than the wavelength of the modes being considered, it
is generally sufficient~\cite{RHBcyclic} to consider only the dominant
modes in each phase (except during the bounce phase and when mode
matching conditions are applied~\cite{Durrer}). In our case however,
the time scale  between the small bounces is very small compared to
the length scales of interest, and hence we cannot just focus on the
dominant modes because for those small time durations the  subdominant
modes can also have an effect on the primordial power spectra.  We
need to  keep all the contributions and give a comprehensive analysis.

This paper is organized as follows. In Sec.~\ref{sec2}, we specify our
setup for the intermediate small bounce feature and discuss about the
relevant scales involved.  In Sec.~\ref{sec3}, we present the
calculation of power spectrum. A detailed presentation of the required
matching conditions is given and the specific results are given for
the two specific small inter bounce features we considered.  An
analysis of these results for the power spectrum is then given in
Sec.~\ref{sec4}.  In Sec.~\ref{sec5}, we give a generalization for the
case of a large number of small bounces. In Sec.~\ref{sec6} we present
our conclusions. An appendix is included to discuss some of the
technical details.

\section{Setup}
\label{sec2}

We will consider a spatially flat Friedmann-Lemaitre-Robertson-Walker
space-time in which the metric is given by
\begin{equation}
ds^2 \, = \, dt^2 - a(t)^2 d{\bf x}^2 \, ,
\end{equation}
where $t$ is physical time, ${\bf x}$ are the comoving spatial
coordinates, and $a(t)$ is the cosmological scale factor. It will be
convenient to use conformal time $\tau$ related to the physical time
via $dt = a(t) d\tau$. We will consider only linear cosmological
perturbations (see, e.g., Ref.~\cite{MFB} for a comprehensive
review). In this case, fluctuations evolve independently in {}Fourier
space. We will label the fluctuation modes in terms of their comoving
wave number $k$.

\begin{figure}[htbp]
\begin{center}
\subfigure[Model 1: Flat plateau.]  {
  \includegraphics[width=0.4\textwidth]{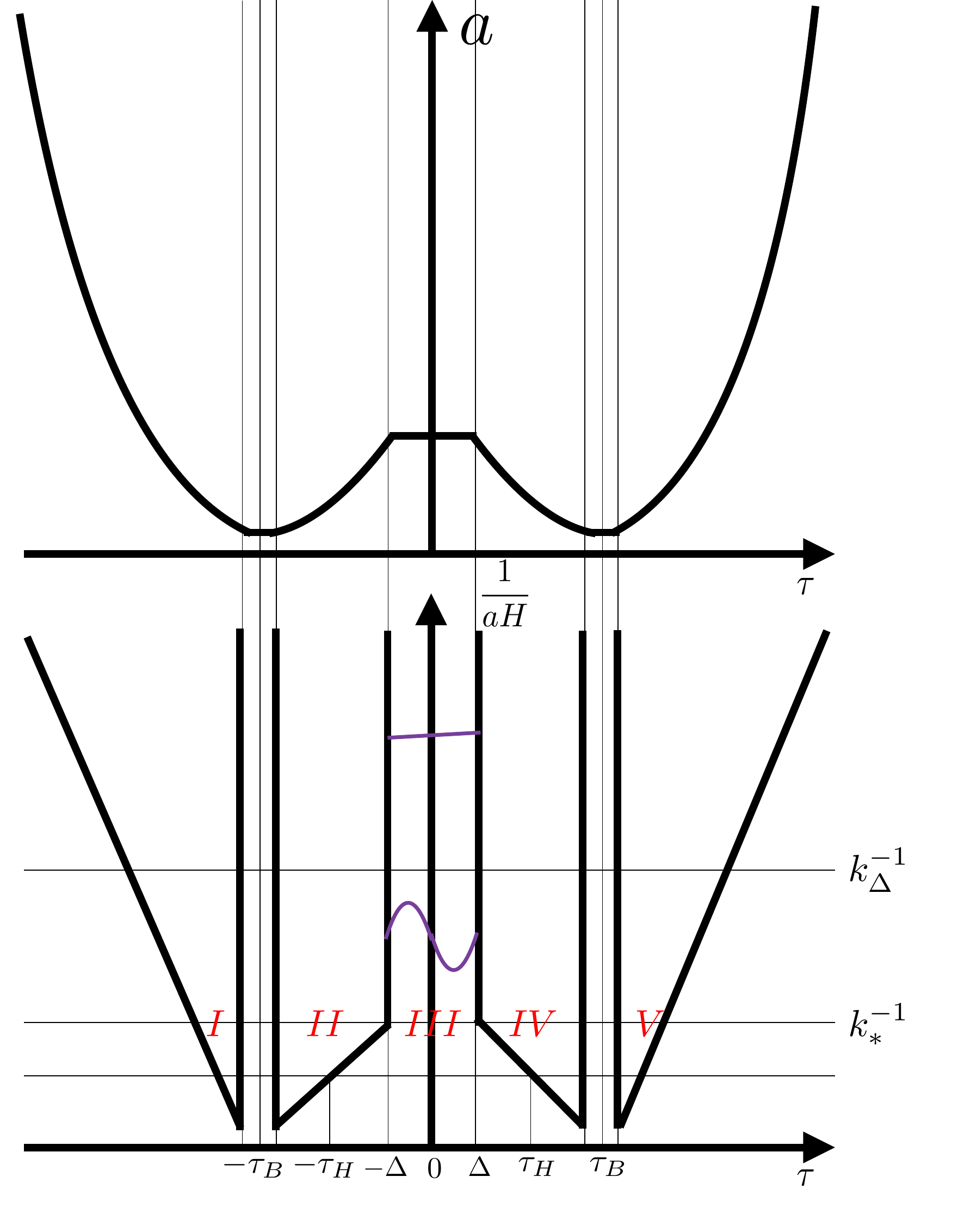}}
\subfigure[Model 2: Instantaneous matching.]{
  \includegraphics[width=0.4\textwidth]{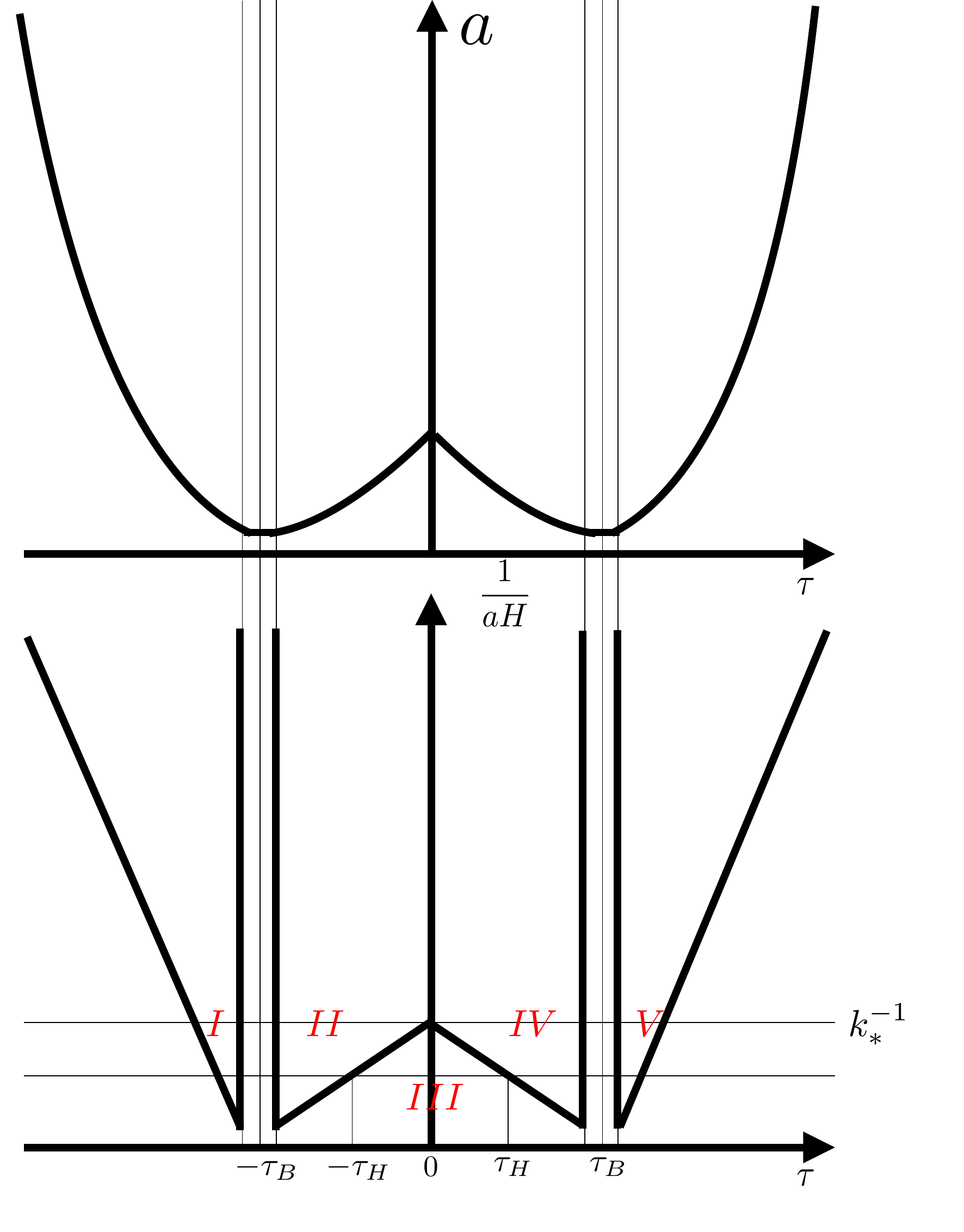}}
\end{center}
\caption{The two upper plots depict the evolution of the scale factor
  as a function of conformal time  $\tau$. The two lower plots
  corresponding to the evolution of the inverse Hubble parameter
  $|aH|^{-1}$ as a function of conformal time. {}For each plot on the
  left hand side, there are 11  characteristic times, from left to
  right, correspond to  $\tau = -\tau_B-\epsilon, -\tau_B,
  -\tau_B+\epsilon, -\tau_H, -\Delta, 0, \Delta, \tau_H,
  \tau_B-\epsilon, \tau_B, \tau_B+\epsilon$.  $-\tau_B$ and $\tau_B$
  are the bouncing points. We made a non-singular bounce by connecting
  $-\tau_B-\epsilon$ and $-\tau_B+\epsilon$, and also
  $\tau_B-\epsilon$ and $\tau_B+\epsilon$.  $-\tau_H(k)$ and
  $\tau_H(k)$ are the times of Hubble radius crossing in Regions II
  and IV,  respectively. $-\Delta$ and $\Delta$ give the duration of
  the interval when the scale factor $a$ is   constant. The two
  magneta curves on the left hand side of the plot are illustrations
  of the  scale of the modes we are considering.  }
\label{bounceplot} 
\end{figure}

We consider a non-singular symmetric bouncing cosmology in which the
cosmological scale factor  $a(\tau)$ has the form shown in
{}Fig.~\ref{bounceplot}, i.e., for which $a(\tau)$ has one
``oscillation'' between the onset and end of the bounce phase.  We
consider the two forms depicted in {}Fig.~\ref{bounceplot}, with one
peak between the time $- \tau_B$ when $a(\tau)$ reaches its  first
minimum, and the time $+ \tau_B$ when the second minimum of $a(\tau)$
is taken on. Specifically, we consider two specific models: Model 1,
with a flat plateau about $\tau = 0$; and the Model 2, with a kink of
$a(\tau)$ at $\tau = 0$, which is the limiting case of the first model
when  the duration of the flat plateau equals to zero. The time
interval of the plateau region is $-\Delta < \tau < \Delta$ given by
some conformal time $\Delta$ with $\Delta < \tau_B$.  The forms shown
in {}Fig.~\ref{bounceplot} are simpler enough such as to allow an
analytical study, but  already of sufficient complexity such as to
provide the main relevant features in the power spectrum  that we
might also observe in some more complex setup for the inter-bounce
features. In particular, our results can be easily generalized to the
case of many oscillations, as we will later discuss in
Sec.~\ref{sec5}.

As seen in {}Fig.~\ref{bounceplot}, for the two forms shown the time
interval can be divided into five intervals. The first is the initial
contracting phase (Phase I) $\tau < - \tau_B$. The second is the
intermediate expanding phase (Phase II). The third is Phase III with
static scale factor, the fourth (Phase IV) is the intermediate
contracting phase and Phase V is the final expanding phase.
{}Fluctuation modes exit the Hubble radius in the initial  contracting
phase. The top panels in {}Fig.~\ref{bounceplot} give a sketch of the
scale factor, the lower panels show the corresponding time evolution
of the comoving Hubble radius. The vertical axis of the lower panels
can also be viewed as a label for comoving wavelength. In this way, it
is easy to read off when various modes enter and exit the Hubble
radius.

All scales re-enter and re-exit the Hubble radius several times since
at the extrema of $a(t)$ the Hubble radius is infinite. We will treat
the transitions at $-\tau_B$ and $\tau_B$ as
instantaneous\footnote{From the point of view of string theory, one
  may view the time interval we are cutting out as the string time
  scale, the time scale where the effective field theory description
  will break down.}. More specifically, we will cut out a time
interval 
\begin{equation}
-\tau_B - \epsilon < \tau < - \tau_B + \epsilon,
\end{equation}
(with $\epsilon \ll \tau_B$) and correspondingly another time interval
of the same length about $\tau_B$ and we will match the solutions
between the neighboring phases making use of the matching conditions
given in Refs.~\cite{HV, DM}, which are the cosmological version of
the Israel~\cite{Israel} ones\footnote{Note that applying these
  matching conditions directly between a contracting phase and an
  expanding phase may be a bit suspect since the background does not
  satisfy the matching conditions (see Ref.~\cite{Durrer} for a
  detailed discussion of this point). However, as long as the matching
  surface is unambiguously determined, the matching conditions for the
  fluctuations can indeed be applied.}. Hence, the only Hubble
re-entry which is important to us is the one which occurs  between $-
\tau_B$ and $+ \tau_B$.

In the first model, given by the plots on the left shown in
{}Fig.~\ref{bounceplot},  there are two characteristic comoving length
scales. The first is  $k_*^{-1}$ which is defined as the length which
re-enters the Hubble radius  at time $- \Delta$. The second one,
$k_{\Delta}^{-1}$, is the mode which undergoes one oscillation between
$\tau = - \Delta$ and $\tau = + \Delta$ (we are assuming
$k_{\Delta}^{-1} >  k_*^{-1}$ {\textemdash} if this is not satisfied
then we recover the results for the second model, given by the plots
on the right  shown in {}Fig.~\ref{bounceplot}).  Modes with
wavelength smaller than $k_*^{-1}$ enter the Hubble radius during
Phase II and exit again during Phase IV. Modes with $k_*^{-1} < k^{-1}
< k_{\Delta}^{-1}$ are inside the Hubble radius only during Phase
III. {}For these modes the matching occurs at times $- \Delta$ and $+
\Delta$. This is also true for modes with $k^{-1} > k_{\Delta}^{-1}$.
These modes, however, undergo a negligible amount of oscillations in
Phase III.  In the case that $k_\Delta^{-1}>k_*^{-1}$, we have three
different  behaviors of the power spectrum. {}For the very large scale
modes $k^{-1}\gg k_\Delta^{-1}$,  the power spectrum does not feel the
influence of the small bump of the scale factor.  {}For the modes
$k_*^{-1}<k^{-1}<k_\Delta^{-1}$, there is a complicated change of the
power spectrum induced by the flat plateau. {}For the modes
$k^{-1}<k_*^{-1}$,  the change of the power spectrum approaches the
well known result for cyclic  cosmologies~\cite{RHBcyclic}, as we will
explicitly verify later on below.  In the case that
$k_\Delta^{-1}<k_*^{-1}$, there is  only one characteristic scale
$k_*^{-1}$. The mode with $k^{-1}>k_*^{-1}$ will not  feel the
influence of the bump, while the mode with $k^{-1}<k_*^{-1}$ will be
changed by the bump according to the well known result for cyclic
cosmologies.  A special situation belonging to this case is the
limiting case $\Delta \rightarrow 0$.

We divide the evolution of fluctuation modes into five regions as
shown on {}Fig.~\ref{bounceplot}
\footnote{Note that the ``regions'' defined here are not the same as
  the ``phases'' defined above. The ``phases'' refer to particular
  behaviors of the scale factor, the ``regions'' to particular
  behaviors of the fluctuation modes. Phases I and V are equal to
  Regions I and V, but for the others there is a difference.}.  The
five regions are denoted by Region I, Region II, Region III, Region IV
and Region V,  respectively. Region I and Region II are separated by
the time $-\tau_B$. Region IV  and Region V are separated by
$\tau_B$. The separation between Regions II and III, and between
Regions III and IV are more complicated. Because of the existence of
the flat  plateau (or of the local maximum of the comoving Hubble
radius in the case of Model 2),  we can see that there is a clear
distinction between the large scale and small scale modes  separated
by a characteristic scale $k_*^{-1}$. Small scale modes
(i.e. $k^{-1}<k_*^{-1}$) enter the Hubble radius at time $-\tau_H(k)
\leq - \Delta$ and exit the Hubble radius at $\tau_H(k) > \Delta$, and
the separation between Regions II and III and  between Regions III and
IV are given by $-\tau_H(k)$ and $\tau_H(k)$, respectively.  For large
scale modes, the separations between Regions II and III, and between
Regions III and IV  are given by the times $-\Delta$ and $\Delta$,
respectively, because the modes enter and exit the Hubble radius at
these two times. The situation in the case of Model 2  is simpler. The
evolution of small scale modes is the same as the case with a flat
plateau,  while the large scale modes have only four regions which we
denote by Regions I,  II,  IV and V, respectively. The separation
between Regions II and IV, in this case,  is the time $\tau= 0$.

\section{The Computation of the Power Spectrum}
\label{sec3}

We are interested in the power spectrum of the primordial curvature
perturbation $\zeta$ (see, e.g., Ref.~\cite{MFB} for a review of the
theory of cosmological perturbations). We quantize the linear
fluctuations and write them in terms of the more convenient
Mukhanov-Sasaki variable $v$.  In the case of a constant equation of
state, the relation between $\zeta$ and $v$ is
\begin{align}
v = C a \zeta \, ,
\end{align}
where $C$ is a constant. Thus, the equation of motion for the mode
function $v$, in momentum  space, is given by
\begin{align}
v'' + \bigg(k^2 - \frac{a''}{a} \bigg) v = 0 \, .
\end{align}
{}For the scale factor $a\sim \tau^q$, the solution of the
Mukhanov-Sasaki equation is given by
\begin{align}
v (\tau) = c_1(k) \sqrt{\tau} J_{\alpha} (k\tau) +  c_2(k) \sqrt{\tau}
Y_{\alpha} (k\tau),\quad \alpha\equiv\ q-\frac{1}{2}  ,
\end{align}
where $J(x)$ and $Y(x)$ are Bessel functions of the first and second
kind, respectively. On sub-Hubble scales, the solutions are oscillatory,
on super-Hubble scales they can be approximated by a power law. To see this, 
we note that the expansion of the Bessel function solutions for small argument,
$x\ll 1$, is
\begin{align}
J_\alpha (x)  &= \sum_{m = 0}^\infty
\frac{(-1)^m}{m!\Gamma(m+\alpha+1)} \left( \frac{x}{2}
\right)^{2m+\alpha}, \\ Y_\alpha (x)  &=
\frac{\cos(\alpha\pi)}{\sin(\alpha\pi)} J_\alpha(x) -
\frac{1}{\sin(\alpha\pi)}J_{-\alpha}(x)  \nonumber \\ &=
\frac{\cos(\alpha \pi)}{\sin(\alpha \pi)} \sum_{m = 0}^\infty
\frac{(-1)^m}{m!\Gamma(m+\alpha+1)} \left( \frac{x}{2}
\right)^{2m+\alpha} - \frac{1}{\sin(\alpha\pi)}\sum_{m = 0}^\infty
\frac{(-1)^m}{m!\Gamma(m-\alpha+1)} \left( \frac{x}{2}
\right)^{2m-\alpha},
\end{align}
we can express the mode function in terms of a series expansion
\begin{align}
v(\tau) = \sum_{m=0}^{\infty} d_{1m} (k) \tau^{q+2m} +
\sum_{m=0}^{\infty} d_{2m} (k) \tau^{1-q+2m}  ,
\label{vtaum}
\end{align}
where $d_{1m}(k)$ and $d_{2m}(k)$ are given, respectively, by 
\begin{align}
d_{1m}(k) & = \left[ c_1(k) + \frac{\cos(\alpha\pi)}{\sin(\alpha\pi)}
  c_2(k) \right]  \frac{(-1)^m}{m!\Gamma(m+\alpha+1)} \left(
\frac{k}{2} \right)^{2m+\alpha}, \\ d_{2m}(k) & =
-\frac{c_2(k)}{\sin(\alpha\pi)} \frac{(-1)^m}{m!\Gamma(m-\alpha+1)}
\left( \frac{k}{2} \right)^{2m-\alpha}  .
\end{align}
Since we are interested in those modes that went classical (crossed
the Hubble radius), such that $k \tau \ll 1$, the higher order terms
of $v(\tau)$ given by $m> 0$ are subleading.   Thus, in the following,
we can just focus on the $m=0$ terms in Eq.~(\ref{vtaum}). 

The scale factors and the solutions to the mode functions of the five
regions can be obtained by  shifting the time coordinate. Thus, they
are given, respectively, by
\begin{itemize}
\item
Region I (contracting), where $\tau<-\tau_B$, we have that
 
\begin{align}
a \sim  (-\tau-\tau_B)^{q_2} \sim(-t-t_B)^{p_2}, \quad v_{1} = c_{11}
(-\tau_B-\tau)^{1-q_2} + c_{12} (-\tau_B-\tau)^{q_2},
\end{align}

\item 
Region II (expanding), where  $ -\tau_B<\tau<-\Delta\,\,\, {\rm
  for}\,\,\, k^{-1}>k_*^{-1},   \,\,\, {\rm
  and}\,\,\,-\tau_B<\tau<-\tau_H \,\,\, {\rm for}\,\,\,
k^{-1}<k_*^{-1}$, we have that

\begin{align}
a\sim (\tau+\tau_B)^{q_1}\sim(t+t_B)^{p_1},\quad v_2 =
c_{21}(\tau+\tau_B)^{1-q_1} + c_{22} (\tau+\tau_B)^{q_1},
\end{align}

\item 
Region III (intermediate), where $-\Delta<\tau<\Delta\,\,\, {\rm
  for}\,\,\, k^{-1}>k_*^{-1}, \,\,\, {\rm and}\,\,\,
-\tau_H<\tau<\tau_H \,\,\, {\rm for}\,\,\, k^{-1}<k_*^{-1}$, we have
that

\begin{align}
a\sim {\rm constant\,\,\,in\,\,\,-\Delta<\tau<\Delta} , \quad v_3 =
c_{31} e^{i k \tau} + c_{32} e^{-i k \tau},
\end{align}

\item 
Region IV (contracting), where $\Delta<\tau<\tau_B\,\,\, {\rm
  for}\,\,\, k^{-1}>k_*^{-1},   \,\,\, {\rm and}\,\,\,
\tau_H<\tau<\tau_B \,\,\, {\rm for}\,\,\, k^{-1}<k_*^{-1}$, we have
that

\begin{align}
a \sim  (-\tau+\tau_B)^{q_1} \sim (-t+t_B)^{p_1},\quad v_4 = c_{41}
(-\tau+\tau_B)^{1-q_1} +c_{42} (-\tau+\tau_B)^{q_1},
\end{align}

\item Region V (expanding), where $\tau>\tau_B$, we have that
 
\begin{align}
a\sim  (\tau-\tau_B)^{q_2} \sim (t-t_B)^{p_2},\quad v_5 = c_{51} (\tau
- \tau_B)^{1-q_2} + c_{52} (\tau-\tau_B)^{q_2}  .
\end{align} 
\end{itemize}
Note that in the above, the time $\tau_H$ depends on $k$,
$\tau_H\equiv \tau_H(k)$. To simplify the notation we do not write the
k-dependence explicitly. Note also that $q_2$ (the power of the scale
factor in conformal time) and $p_2$ (the power of the scale factor in
cosmological time) are the indices of the scale factor during the
initial contracting and final expanding phase, and that $q_1$ and
$p_1$ are the indices in the intervening periods. In particular, the
index 1 in these quantities should not be confused with the index in
Region I.  Note also that the $q_i$ are related to the $p_i$ through
the relation

\begin{equation}
q_i = \frac{p_i}{1-p_i},\;\;\;i=1,2.
\label{qipi}
\end{equation}
\subsection{Matching conditions}

Let us now discuss the process of matching between the different
regions discussed above.  The matching conditions for the metric
across a space-like hypersurface were derived  in
Hwang-Vishniac~\cite{HV} and Deruelle-Mukhanov~\cite{DM} and are the
generalization of the Israel matching conditions \cite{Israel}. For
cosmological fluctuations, the matching  conditions say the solutions
in different regions can be connected by enforcing two conditions,
namely the continuity of both $v$ and its derivative across the
boundary surface. 

As mentioned earlier, for the two non-singular bouncing points $-
\tau_B$ and $\tau_B$  we match the solutions at times
$\mp\tau_B-\epsilon$ and $\mp\tau_B+\epsilon$, neglecting any
evolution in the intervening time period. This is similar to what was
done in Refs.~\cite{Elisa, Subodh}.  A second justification of this
method (in addition to the one given earlier) is that for modes we are
interested in,  the time interval $2\epsilon$ is so small, thus the
mode functions do not  have enough time to oscillate inside the Hubble
radius. On the other hand, in our first model (with a flat plateau for
$a(t)$, model 1) we consider the case that the interval  $2\Delta$ is
sufficiently long such that some of the modes we are interested in
have time to  oscillate while the mode is inside the Hubble
radius. Very large scale modes, on the other hand, still do not
oscillate inside the Hubble radius.

\subsubsection{Between region I and region II}

The matching conditions are
\begin{align}
v_1(-\tau_B-\epsilon) = v_2 (-\tau_B+\epsilon), \quad
v'_1(-\tau_B-\epsilon) = v'_2 (-\tau_B+\epsilon) \, .
\end{align}
We write down the equations explicitly in terms of coefficients
$c_{ij}$ of the  fundamental solutions of the equation of motion. The
index $i$ stands for the region, the index $j$ (either $1$ or $2$)
running over the two different modes:

\begin{align}
\begin{pmatrix}
\epsilon^{1-q_2} & \epsilon^{q_2} \\ -(1-q_2)  \epsilon^{-q_2} &
-q_2\epsilon^{q_2-1}
\end{pmatrix} \begin{pmatrix}
c_{11} \\ c_{12}
\end{pmatrix} = \begin{pmatrix}
\epsilon^{1-q_1} & \epsilon^{q_1} \\ (1-q_1) \epsilon^{-q_1} & q_1
\epsilon^{q_1-1}
\end{pmatrix} \begin{pmatrix}
c_{21} \\ c_{22}
\end{pmatrix} .
\end{align}

\subsubsection{Between region II and region III}

To be explicit, we focus in this case on the large scale modes for
which  the time duration of Region III is $-\Delta<\tau<\Delta$. {}For
the small scale modes which enter the Hubble radius before $- \Delta$,
we just make the substitution  $\Delta\rightarrow\tau_H(k)$. Apart
from that the discussion is the same.  {}For the next subsection the
convention will be the same.  The matching conditions in this case are
\begin{align}
v_2(-\Delta) = v_3 (-\Delta),\quad  v'_2(-\Delta) = v'_3 (-\Delta),
\end{align}
which in matrix form can be expressed as

\begin{align}
\begin{pmatrix}
\delta^{1-q_1} & \delta^{q_1} \\ (1-q_1) \delta^{-q_1} & q_1
\delta^{q_1-1} 
\end{pmatrix} \begin{pmatrix}
c_{21} \\ c_{22}
\end{pmatrix} = \begin{pmatrix}
e^{-ik \Delta} & e^{i k \Delta} \\ i k e^{-ik\Delta} & - i k e^{i k
  \Delta}
\end{pmatrix} \begin{pmatrix}
c_{31} \\ c_{32}
\end{pmatrix},
\end{align}
and where in the above equation we have defined $\delta$ as

\begin{equation}
\delta = \tau_B - \Delta .
\label{delta}
\end{equation}
Note that for the small scale modes the definition of $\delta$ should
be changed to  $\delta\to \delta(k) \equiv \tau_B-\tau_H(k)$.

\subsubsection{Between region III and region IV}

The matching conditions are
\begin{align}
v_3(\Delta) = v_4 (\Delta), \quad v'_3(\Delta) = v'_4 (\Delta) \, .
\end{align}
In matrix form this yields
\begin{align}
\begin{pmatrix} 
e^{ik\Delta} & e^{-ik\Delta} \\ i k e^{ik\Delta} & -i k e^{-ik\Delta}  
\end{pmatrix} \begin{pmatrix}
c_{31} \\ c_{32}
\end{pmatrix} = \begin{pmatrix}
\delta^{1-q_1} & \delta^{q_1} \\ -(1-q_1) \delta^{-q_1} &
-q_1\delta^{q_1-1}
\end{pmatrix} \begin{pmatrix}
c_{41} \\ c_{42}
\end{pmatrix}.
\end{align}

\subsubsection{Between region IV and region V}

The matching conditions in this case are
\begin{align}
v_4(\tau_B-\epsilon) = v_5 (\tau_B+\epsilon),\quad
v'_4(\tau_B-\epsilon) = v'_5 (\tau_B+\epsilon) \, .
\end{align}
In matrix form this yields
\begin{align}
\begin{pmatrix}
\epsilon^{1-q_1} & \epsilon^{q_1} \\ -(1-q_1) \epsilon^{-q_1} & -q_1
\epsilon^{q_1-1}
\end{pmatrix} \begin{pmatrix}
c_{41} \\ c_{42}
\end{pmatrix} = \begin{pmatrix}
\epsilon^{1-q_2} & \epsilon^{q_2} \\ (1-q_2) \epsilon^{-q_2} & q_2
\epsilon^{q_2-1}
\end{pmatrix} \begin{pmatrix}
c_{51} \\ c_{52}
\end{pmatrix}.
\end{align}

\section{Analysis and Results for the Power Spectrum}
\label{sec4}

Combining the results of the previous section we find that the final
mode coefficients can be written in terms of the initial ones via
\begin{align}\nonumber
& \mathcal C_5 = 
\begin{pmatrix}
c_{51} \\ c_{52}
\end{pmatrix},
\end{align}
where\footnote{Where we are neglecting the coefficient $c_{12}$ of 
the decaying mode in the initial phase.} 

\begin{eqnarray}
c_{51}&=&\frac{c_{11}}{(1-2q_2)(1-2q_1)2k(1-2q_1)} \left[ a_{11} (1 -
  2 (q_1-1) q_1 - 2 (q_2 -1 ) q_2) \right.  \nonumber \\ &-&
  \left. a_{12} ( (q_1 - q_2 - 1) (q_1 + q_2 - 2) \epsilon^{1 - 2 q_1}
  ) + a_{21} (q_1 - q_2 + 1) (q_1 + q_2) \epsilon^{2 q_1 - 1} \right],
\label{c51}
\\ c_{52}&=&\frac{c_{11}}{(1-2q_2)(1-2q_1)2k(1-2q_1)} \left[ 2 a_{11}
  (1 + q_1 - q_2) (-2 + q_1 + q_2) \epsilon^{1 - 2 q_2} \right.
  \nonumber \\ &+& \left. a_{12} (-2 + q_1 + q_2)^2 \epsilon^{-2 (-1 +
    q_1 + q_2)} + a_{21} (1 + q_1 - q_2)^2 \epsilon^{2 (q_1 -
    q_2)}\right],
\label{c52}
\end{eqnarray}
and
\begin{align}\label{a11}
a_{11} & = \frac{2\left[k^2\delta^2+(q_1-1)q_1\right]\sin
  (2k\Delta)}{\delta} - 2 k \cos (2 k \Delta), \\
\label{a12}
a_{12} & = \delta^{2q_1-2}
\left[-4kq_1\delta\cos(2k\Delta)-2(q_1-k\delta)(q_1+k\delta)
  \sin(2k\Delta)\right],
\\
\label{a21}
a_{21} & =
\delta^{-2q_1}\left[2(q_1-k\delta-1)(q_1+k\delta-1)\sin(2k\Delta)-
  4k(q_1-1)\delta\cos(2k\Delta)\right]
.
\end{align}
Note that these coefficients oscillate as a function of $k$. These
oscillations are important, however, only for small wavelength
fluctuations. {}For these we will obtain oscillations in the power
spectrum.  The final general result for the power spectrum is given by 
\begin{align} 
P_{\zeta} & = \zeta^2 k^{3}  \sim \left( \frac{v}{z} \right)^2 k^{3}
\nonumber  \\ & = \left[
  \frac{c_{51}(\tau-\tau_B)^{1-q_2}+c_{52}(\tau-\tau_B)^{q_2}}{(\tau-\tau_B)^{q_2}}
  \right]^2 k^{3}    \nonumber \\  &= \left[ c_{51}
  (\tau-\tau_B)^{1-2q_2} + c_{52} \right]^2 k^{3} \, .
\label{finalPzeta}
\end{align}
Below we will analyze some of the specific cases given by our two
models when applying the result given by Eq.~(\ref{finalPzeta}).

\subsection{Limiting case of Instantaneous matching}

We first consider the limit as the duration of the plateau region of
$a(t)$ goes to zero, corresponding to what we have denoted by Model 2
in Sec.~\ref{sec2}. This is the limit $\Delta\rightarrow 0$. In this
case, large scale modes $k^{-1} > k_*^{-1}$ do not enter the Hubble
radius in the region near $t = 0$, and we can set $\Delta = 0$ in the
matching condition equations, i.e.,

\begin{align}
\sin(2k\Delta) \rightarrow 0, \quad \cos(2 k \Delta) \rightarrow 1 \,
.
\end{align}
On the other hand, small scale modes $k^{-1} < k_*^{-1}$ will enter
the Hubble radius at a time given by $-\tau_H(k)$, and in the matching
condition equations we must replace $\Delta$ by  $\tau_H(k)$.

\subsubsection{Large scale modes $k^{-1} > k_*^{-1}$}

Let us first consider the case for large scale modes $k^{-1} >
k_*^{-1}$. In this case we have that
\begin{align}
a_{11}\rightarrow -2k, \quad a_{12}\rightarrow -4k q_1\delta^{2q_1-1},
\quad a_{21}\rightarrow -4k (q_1-1)\delta^{-2q_1+1}.
\end{align}
Because we are interested in the parameter region $1/3<p<1$, then,
written in terms of $q$,  we have $q>1/2$. So the $c_{51}$ mode in the
expression for the power spectrum Eq.~(\ref{finalPzeta}) is a decaying
solution. Hence, we can focus on the constant mode $c_{52}$, and thus
the power spectrum in this case becomes 
\begin{align} \nonumber
&P_{\zeta} \sim c_{52}^2 k^{3} =\left\{
  \frac{c_{11}}{(1-2q_2)(1-2q_1)2(1-2q_1)}  \right. \\ \nonumber &
  \left.  \times \left[ -4 (1 + q_1 - q_2) (-2 + q_1 + q_2)
    \epsilon^{1 - 2 q_2}-4q_1\delta^{2q_1-1}  (-2 + q_1 + q_2)^2
    \epsilon^{-2 (-1 + q_1 + q_2)} \right. \right. \nonumber \\   &
    \left. \left. -4(q_1-1)\delta^{-2q_1+1} (1 + q_1 - q_2)^2
    \epsilon^{2 (q_1 - q_2) } \right] \right\}^2 k^{3}  .
\end{align}

The initial power spectrum is 
\begin{align}
P_i = P_{\zeta} (-\tau_B-\epsilon) = \zeta^2 k^{3}  = c_{11}^2
\epsilon^{2-4q_2}  k^{3}
\end{align}
and, thus, we can relate the final to the initial power spectrum as
\begin{align}\label{powerspectrum}
P_{\zeta} = \left(A_1  + A_2   \delta^{2q_1-1} \epsilon^{1-2 q_1 } +
A_3\delta^{-2q_1+1} \epsilon^{2q_1-1} \right)^2 P_i \, ,
\end{align}
where $A_1$, $A_2$ and $A_3$ are constants that do not depend on
$k$. Their explicit forms are 
\begin{align} 
&A_1 = \frac{-2(1+q_1-q_2)(-2+q_1+q_2)}{(1-2q_2)(1-2q_1)^2},\\  &A_2 =
  \frac{-2q_1(-2+q_1+q_2)^2}{(1-2q_2)(1-2q_1)^2}, \\ &A_3 =
  \frac{-2(q_1-1)(1+q_1-q_2)^2}{(1-2q_2)(1-2q_1)^2}.
\end{align}

{}For very large scale modes $k^{-1} \gg  k_*^{-1}$, $\delta \to
\tau_B$ and $\delta$ can be approximated as a constant time
interval. Thus, the power spectrum in this case becomes

\begin{align}\label{powerspectrum1}
(A_1  + A_2  \tau_B^{2q_1-1} \epsilon^{1-2 q_1 } + A_3
  \tau_B^{-2q_1+1} \epsilon^{2q_1-1} )^2 P_i .
\end{align}
The first conclusion we draw from this result is that the shape of the
spectrum for large scale modes does not change during the bounce. This
agrees with the conclusions of previous work on simple bounce
models~\cite{matterBounce-2}. The amplitude, on the other hand, is
amplified. Recall that $2q_i - 1 > 0$, and that $\epsilon \ll \tau_B$.
Hence, it is the second term in Eq.~(\ref{powerspectrum1}) which
dominates, and we conclude that the amplitude of the spectrum is
amplified by a factor of
\begin{equation}
{\cal A} \, = \, A_2^2 \bigg(\frac{\tau_B}{\epsilon}\bigg)^{4q_1 - 2} .
\label{Amplitude}
\end{equation}
This result can also be understood easily: {}Fluctuations grow both in
the contracting and in the expanding phase. In fact, the fluctuations
diverge in the limit when the scale factor becomes zero. Hence,
without an effective cutoff $\epsilon$ we would get a divergence in
the spectrum. With a cutoff, the enhancement factor of the amplitude
of the power spectrum will be determined by the dimensionless ratio
between $\tau_B$ and $\epsilon$ to a power which depends on the growth
rate of the fluctuations on super-Hubble scales, i.e., on $q_1$ (see
the discussion of these issues in a more general context in the review
article Ref.~\cite{RHBbounceReview}).

\subsubsection{Small scale modes $k^{-1} < k_*^{-1}$}

{}For small scale modes $k^{-1}<k_{*}^{-1}$, we set $\delta$ equal to
the Hubble  crossing time. Thus, we can use the Hubble crossing
condition $a H= k$, which from $a\sim \tau^q$ gives
\begin{align}
\delta =q_1  k^{-1} .
\end{align}
But we need to have $k \tau_H = k\tau_B - k\delta= k\tau_B - q_1 $.
As a consequence of the oscillations in the coefficients $a_{ij}$,
Eqs.~(\ref{a11}), (\ref{a12}) and (\ref{a21}), the final power
spectrum of fluctuations will oscillate for small wavelengths. This is
explicitly manifested when we show a numerical example for the power
spectrum in {}Fig.~\ref{bounceplot1},  where we chose an initial
pre-bounce spectrum which is scale-invariant. We see that the
scale-invariance of the spectrum is maintained on large scales, but
that on small length scales there is both a change in the slope of the
spectrum, and superimposed oscillations.

In the following we discuss in what range we can reproduce the results
of Ref.~\cite{RHBcyclic}, which hold for a cyclic cosmology. In that
work, it was found that for modes which re-enter the Hubble radius
during the bounce phase, the index of the spectrum of cosmological
perturbations changes during each cycle. {}For a matter-dominated
contracting phase the change in the index $n_s$ of the power spectrum
was determined to be $\Delta n_s = - 2$. 

The results of Ref.~\cite{RHBcyclic} are applicable when
$k^{-1}<k_*^{-1}$, but for quite large scales such that
$k^{-1}\rightarrow k_*^{-1}$. In this range we have
$k\tau_H\rightarrow 0$.  Using this in Eq.~(\ref{powerspectrum}), we
obtain that
\begin{align}\label{powerspectrum_cyclic}
P_f \, = \,  \left[ A_1  + A_2 (k \epsilon/q_1)^{1-2q_1} +  A_3
  (k\epsilon/q_1)^{2q_1-1} \right]^2 P_i.
\end{align}
Since $2q_1 - 1 > 0$ it is the second term in
Eq.~(\ref{powerspectrum_cyclic}) which dominates.  Hence, we conclude
that there is a change in the index of the power spectrum by 
\begin{equation}
\Delta n_s = 2-4 q_1= - 2 \frac{3 p_1 - 1}{1 - p_1} ,
\label{tiltns}
\end{equation}
which coincides with the results of Ref.~\cite{RHBcyclic}.  This is as
expected because the case studied in Ref.~\cite{RHBcyclic} corresponds
to a big bounce where $\delta$ is (cosmologically) large.

\subsection{Case with a flat plateau}

In the case with a flat plateau and when $\Delta$ is very small, we
have just one  characteristic comoving mometum. However, when $\Delta$
is big, we have two  key comoving momenta which are characterized by
the mode which cross the Hubble radius at $\Delta$ and $\tau_B-
\Delta$, respectively. In this subsection, we  would like to analyze
in detail these two cases.

{}First we would like to calculate the critical comoving momentum
$k_*^{-1}$.  We start by analyzing the Hubble parameter $H$.  The
corresponding comoving Hubble parameter in region II is

\begin{align}
aH =q_1 (\tau+\tau_B)^{-1}.
\end{align}
The critical scale $k_{*}$, which is obtained by
$k_*=aH(\tau=-\Delta)$, is therefore
\begin{align}
k_{*} =  q_1 (\tau_B - \Delta)^{-1}.
\end{align}
The analysis here is similar to the instantaneous matching case of the
previous subsection and we can obtain the power spectrum as
\begin{align} 
&P_{\zeta} \sim c_{52}^2 k^{3} = \left\{ \frac{ 1
  }{(1-2q_2)(1-2q_1)2k(1-2q_1)}  \right.  \nonumber \\ & \left. \times
  \left[  2 a_{11} (1 + q_1 - q_2) (-2 + q_1 + q_2)   + a_{12} (-2 +
    q_1 + q_2)^2 \epsilon^{1 -2 q_1 } \right. \right.  \nonumber \\ &
    \left. \left. + a_{21} (1 + q_1 - q_2)^2 \epsilon^{2 q_1 -1}
    \right] \right\}^2 P_i ,
\label{powerspectrumforinstantaneous}
\end{align}
where $a_{11}$, $a_{12}$ and $a_{21}$ were already defined by
Eqs.~\eqref{a11}, \eqref{a12} and \eqref{a21}, respectively.

\subsection{Model with no Region III}

Let us here consider the model with no Region III. We expect that  the
result we obtained in the previous Subsection will approach the result
derived here in the limit when $\Delta\rightarrow 0$.  The matching
condition of Region I and II, Region IV and V are completely the same
as in the flat plateau case,  so here we only write down the matching
condition between Region II and IV:

\begin{align}
v_2(0) = v_4 (0),\quad v'_2(0) = v'_4 (0) ,
\end{align} 
which can be written in terms of the more convenient matrix form
\begin{align}
\begin{pmatrix}
\tau_B^{1-q_1} & \tau_B^{q_1} \\ (1-q_1) \tau_B^{-q_1} & q_1
\tau_B^{q_1-1} 
\end{pmatrix} \begin{pmatrix}
c_{21} \\ c_{22}
\end{pmatrix} = \begin{pmatrix} 
\tau_B^{1-q_1} & \tau_B^{q_1} \\ -(1-q_1) \tau_B^{-q_1} & -q_1
\tau_B^{q_1-1}
\end{pmatrix} \begin{pmatrix}
c_{41} \\ c_{42}
\end{pmatrix}.
\end{align}
Combining these matching results we obtain the power spectrum
completely the same as that of the instantaneous matching of the
previous section.

\subsection{Numerical Examples}

\begin{figure}[htbp]
\begin{center}
\subfigure[The power spectrum as a function of comoving wavenumber
  $k$.]  { \includegraphics[width=0.43\textwidth]{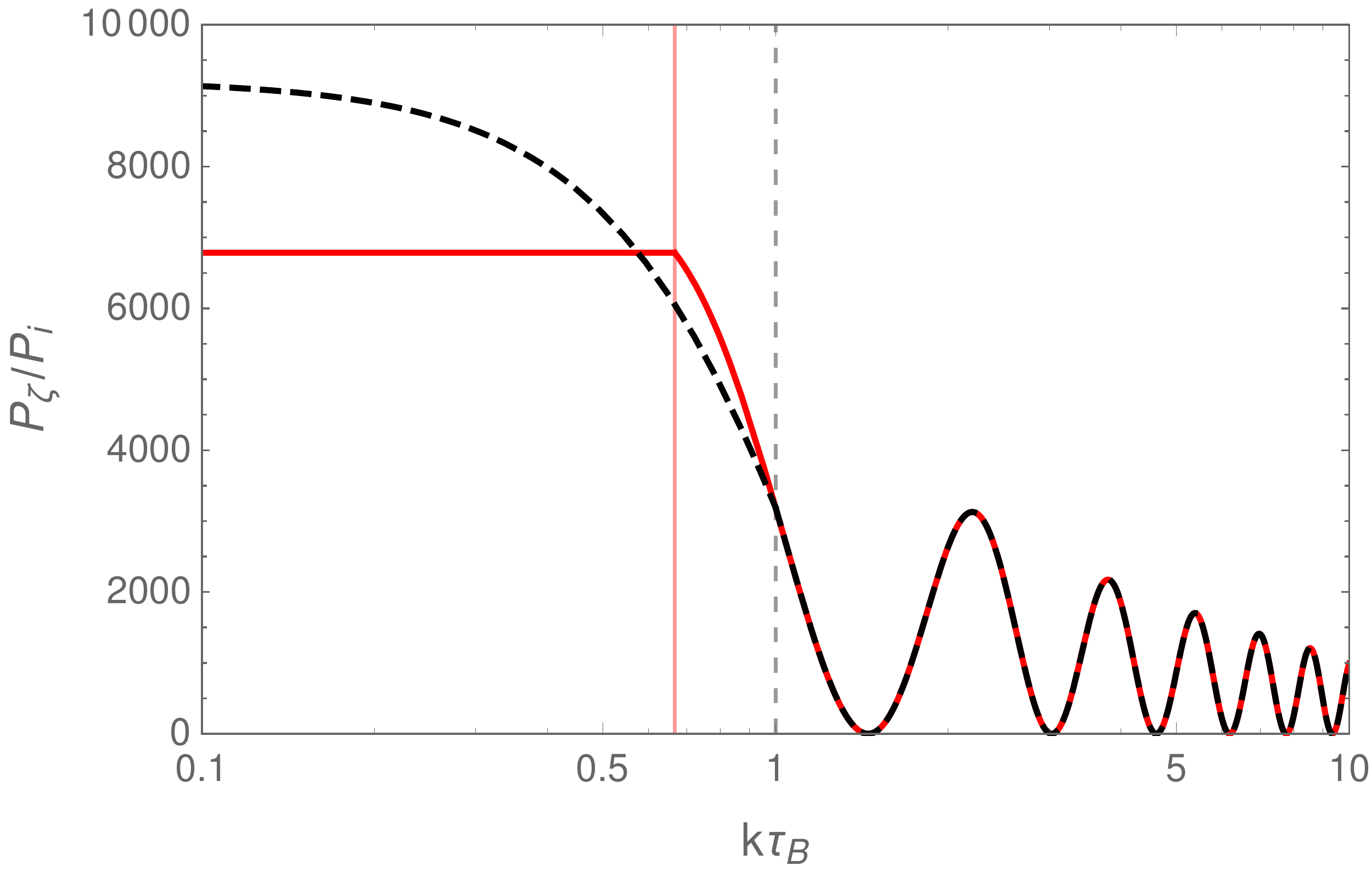}}
\subfigure[The spectral tilt $n_s$ as a function of comoving
  wavenumber $k$.]{
  \includegraphics[width=0.4\textwidth]{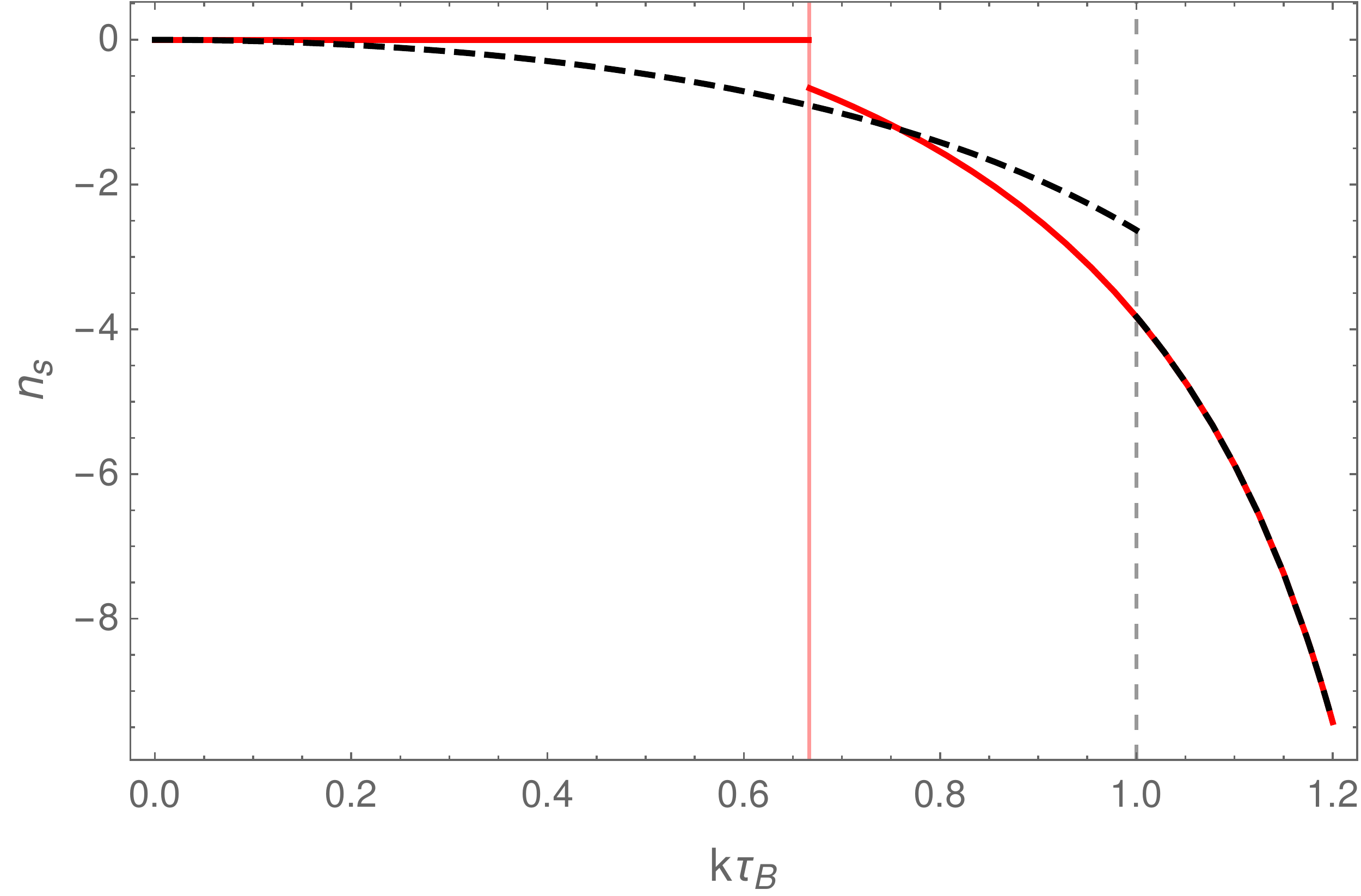}}
\end{center}
\caption{The normalized power spectrum (a) and the spectral tilt (b)
  as a function of the comoving wavenumber times the bounce time
  $\tau_B$ for the parameters values  $q_1= q_2 = 2/3$, $\epsilon =
  0.01 \tau_B$. The black dashed line corresponds to the  case with a flat
  plateau with $\Delta = \tau_B/3$ (Model 1). The red solid line
  corresponds to the case  without a flat plateau $\Delta = 0$ (Model
  2). The solid red vertical line on the left denotes the critical
  comoving momentum  $k_{*,{\rm flat}}$ for the flat plateau case. The
  black dashed vertical line on the right denotes the critical
  comoving momentum  $k_{*,{\rm inst}}$ for the instantaneous kink
  like plateau case.  }
\label{bounceplot1} 
\end{figure}

In {}Fig.~\ref{bounceplot1} we show the form of the spectrum of
cosmological perturbations and its tilt as a function of the comoving
wavenumber $k$ for the two models we have considered in the previous
sections. 

We can clearly identify in {}Fig.~\ref{bounceplot1}(a) the
characteristic scales for each of the two models we have defined in
Sec.~\ref{sec2}. {}For the flat plateau model (Model 1), there are two
relevant comoving scales,

\begin{eqnarray}
&&k_{*,{\rm flat}}= \frac{q_1}{\tau_B-\Delta},
\label{kflat}
\\ &&k_{*,{\rm osc}}= \frac{q_1}{\tau_B}.
\label{kosc}
\end{eqnarray}
In model 1 the spectrum is always evolving. On large scales  $k <
k_{*,{\rm osc}}$ there is both an amplification of the spectrum and a
damping evolution. On small scales $k > k_{*,{\rm osc}}$ the power
spectrum shows superimposed damped oscillations.

In the instantaneous case $\Delta=0$ (Model 2), the characteristic
comoving scale is $k_{*,{\rm inst}} \equiv k_{*,{\rm osc}}$, the same
as Eq.~(\ref{kosc}).  In the model 2, on large scales $k < k_{*,{\rm
    inst}}$ the  spectral shape is unchanged during the bounce and
only the amplitude increases, as identified in Eq.~(\ref{Amplitude}).
On smaller scales $k > k_{*,{\rm inst}}$ there is a change in the
spectral index and  the power spectrum, as in the case of model 1,
shows superimposed damped oscillations.

The results {}Fig.~\ref{bounceplot1}(b) show that for Model 1 (black
dashed line) the spectral tilt always decreases with the momentum. The
discontinuity at  $k_{*,{\rm flat}}$ (denoted by the black dashed
vertical line) is an unphysical feature that appears as a consequence
of the shape we have considered and should not appear in realistic
smooth shapes. The same is true for the Model 2 case (red solid line),
where the discontinuity happens at the characteristic scale $k_{*,{\rm
    inst}}$ in this case and comes from the kink like shape considered
in this model. Other than that,  the spectral index is unchanged (and
null) for large scales modes $k< k_{*,{\rm inst}}$ and then decreases
for  small scale modes $k> k_{*,{\rm inst}}$ and agrees with that of
model 1 from this point on, where the index of the power spectrum for
both models acquires a large red tilt, and there are superimposed
oscillations. 

\begin{figure}[htbp] 
  \centering \includegraphics[width=0.45\textwidth]{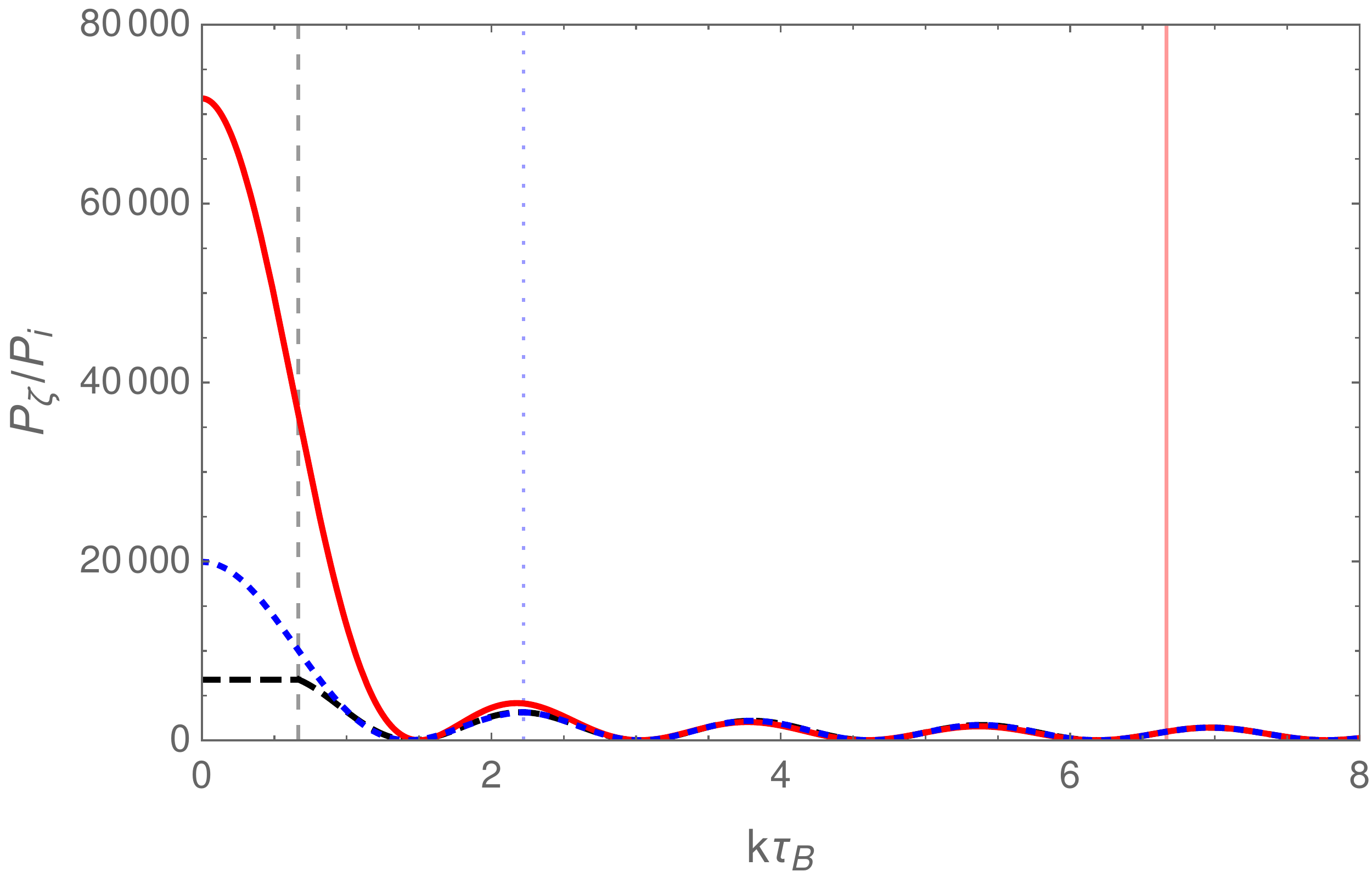} 
  \caption{\label{bounceplot2}  The power spectrum for different
    parameters, namely  $q_1= q_2 = 2/3$, $\epsilon = 0.01 \tau_B$, for the
    cases of $\Delta=0.7 \tau_B$ (blue dotted line), and  $\Delta=0.9
    \tau_B$ (red solid line) for the case of the flat plateau (Model
    1) and for the instantaneous case (Model 2), where $\Delta=0$
    (black dashed line).  Again, the vertical lines denote the
    positions of the characteristic scales $k_*$ for each model.} 
\end{figure}

{}Figure~\ref{bounceplot2} shows the results for different parameter
values, parameters for which the two scales $k_*$ for the models with
and without a plateau for $a(t)$ are more widely separated than they
are for the earlier parameter values. {}For both models there are
oscillations of the power spectrum for $k$ values between the two
critical $k_*$ values.  These results in particular show that as
$\Delta \to \tau_B$, the scale $k_{*,{\rm flat}}$ can occur deeper in
the   oscillating regime $k>k_{*,{\rm osc}}$ for the spectrum. 

A study of the oscillating regime for small scales $k>k_{*,{\rm
    osc}}$, and which is common for both models considered here, is
given in the Appendix~\ref{appA}.  In particular, it is shown that the
envelope function of the power spectrum for the small scale modes
keeps the spectral tilt $n_s = -4q_1+2$, as also seen in the previous
Eq.~(\ref{tiltns}).

\section{Generalization to $n$ small bounces}
\label{sec5}

\begin{figure}[htbp] 
  \centering \includegraphics[width=0.45\textwidth]{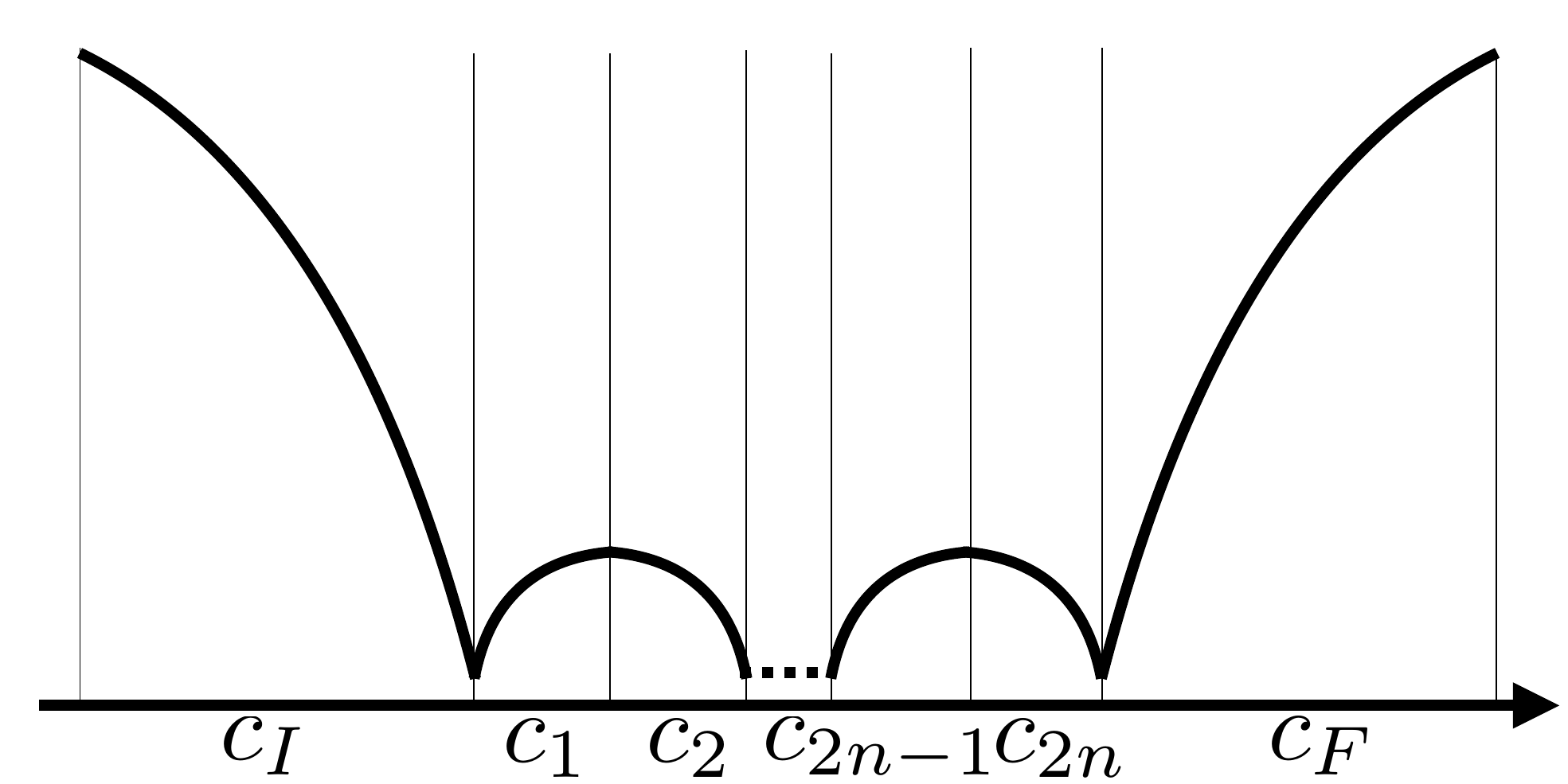}   
    \caption{\label{nbounceplot}  An illustration of the
      generalization to n-vibrations (or small bounces). } 
\end{figure}

In this section, we would like to analyze the case where there are $n$
small bounces {\textemdash} see {}Fig~\ref{nbounceplot}.  Since the
transfer matrix of the flat plateau case is  quite involved, we would
like to first  consider Model 2 (no plateau interval) for illustrative
purposes.

Based on our previous calculations, we can easily write down the
transfer matrices for the  coefficient vector. We define the following
useful matrices
\begin{align}
M_1(\tau,q)  = \begin{pmatrix} \tau^{1-q} & \tau^q \\ -(1-q) \tau^{-q}
  & -q \tau^{q-1}
\end{pmatrix},\quad
M_2(\tau,q)  = \begin{pmatrix} \tau^{1-q} & \tau^q \\ (1-q) \tau^{-q}
  & q \tau^{q-1}
\end{pmatrix} .
\end{align}

{}For large scale modes, we define the combination of matrices
\begin{align}
N =
M^{-1}_2(\tau_B,q_1)M_1(\tau_B,q_1)M^{-1}_2(\epsilon,q_1)M_1(\epsilon,q_1)
,
\end{align}
which becomes
\begin{align}
N = \frac{1}{(1-2q_1)^2} \begin{pmatrix} 4(q_1-1)q_1
  \tau_B^{2q_1-1}\epsilon^{1-2q_1} + 1 &
  2q_1(\epsilon^{2q_1-1}-\tau_B^{2q_1-1})
  \\ 2(q_1-1)(\tau_B^{1-2q_1}-\epsilon^{1-21_1}) &
  4(q_1-1)q_1\epsilon^{2q_1-1}\tau_B^{1-2q_1}+1 
\end{pmatrix}.
\end{align}

Taking the two bump model as an example, we obtain the final
coefficient vector to be
\begin{align}
\mathcal C_F = M_2^{-1} (\epsilon ,q_2) M_1(\epsilon ,q_1) N M_2^{-1}
(\tau_B, q_1) M_1(\tau_B, q_1) M_2^{-1} (\epsilon,q_1) M_1(\epsilon,
q_2) \mathcal C_I  .
\end{align}
We set the initial coefficient matrix $\mathcal C_I$ to be
\begin{align}
\mathcal C_I = \begin{pmatrix} c_{11} \\ 0
\end{pmatrix},
\end{align}
and then we get
\begin{align}\nonumber
c_{52} =&\frac{1}{(1-2q_1)(1-2q_2^3)} \left[ 8  c_{11}
  \left(q_1-1\right){}^2 q_1   \left(q_1-q_2+1\right){}^2   \epsilon
  ^{4 q_1-2 q_2-1}   \tau _B^{2-4 q_1} \right.  \\ \nonumber &
  \left. -4 c_{11}  \left(q_1-1\right)   \left(q_1-q_2+1\right)
  \left(2 q_1^2+2 q_2 q_1-5   q_1+q_2-1\right) \epsilon   ^{2 q_1-2
    q_2} \tau _B^{1-2   q_1} \right.  \\ \nonumber & \left. -4 c_{11}
  q_1   \left(q_1+q_2-2\right)   \left(2 q_1^2-2 q_2   q_1+q_1+3
  q_2-4\right)   \epsilon ^{-2 q_1-2 q_2+2}   \tau _B^{2 q_1-1}
  \right.  \\ \nonumber & \left. -8 c_{11}   \left(q_1-1\right) q_1^2
  \left(q_1+q_2-2\right){}^2   \epsilon ^{-4 q_1-2 q_2+3}    \tau
  _B^{4 q_1-2} \right.  \\ \nonumber & \left. +2 c_{11}   \left(4
  q_1^4-8 q_1^3-4   q_2^2 q_1^2+16 q_2 q_1^2-10  q_1^2+4 q_2^2 q_1-16
  q_2   q_1+14 q_1+2 q_2^2-5   q_2+3\right) \epsilon ^{1-2   q_2}
  \right] .
\end{align}
On very large scale there is no change in the spectral slope, as
expected.

Now we want to deal with the small scale case. We need to define two
more matrices
\begin{align}
L_1 = \begin{pmatrix} e^{-ik\tau_H} & e^{i k \tau_H} \\ ik e^{-i k
    \tau_H}  & -i k e^{i k \tau_H}
\end{pmatrix}, \quad
L_2 = \begin{pmatrix} e^{ik\tau_H} & e^{-i k \tau_H} \\ ik e^{i k
    \tau_H}  & -i k e^{-i k \tau_H}
\end{pmatrix}.
\end{align}
Then we have 
\begin{align}\nonumber
\mathcal C_F = & M_2^{-1} (\epsilon,q_2) M_1(\epsilon,q_1) M_1^{-1}
(\delta,q_1) L_2 L_1^{-1} M_2(\delta,q_1) M_2^{-1} (\epsilon,q_1)
\\ &\times M_1(\epsilon,q_1) M_1^{-1} (\delta,q_1) L_2 L_1^{-1}
M_2(\delta,q_1) M_2^{-1} (\epsilon,q_1) M_1(\epsilon,q_2) \mathcal
C_I,
\end{align}
and the general result has the form 
\begin{align}
P_{\zeta} = \left[\# + \# (\epsilon k)^{2-4q_1} + \# (\epsilon
  k)^{4q_1-2} + \# (\epsilon k)^{1-2q_1}  + \# (\epsilon
  k)^{2q_1-1}\right]^2 P_i \, .
\end{align}
Since we are interested in modes which exit the Hubble radius before
the time $-(\tau_B + \epsilon)$, we consider values of $k$ for which
$\epsilon k \ll 1$. Hence, in this range of $k$ values it is the
second term above which dominates and we find the scaling
\begin{equation}
P_{\zeta} \, \sim \, (\epsilon k)^{4 - 8q_1} P_i  ~.
\end{equation}
Thus, small scale modes acquire a red tilt compared to the initial
spectrum. If the initial spectrum is scale invariant, then the
resulting spectral index for small scale modes is
\begin{equation}
n_s - 1  =  4 - 8q_1  ~.
\end{equation}
Similarly, we can obtain the spectral index change for $n$ small
bounces, which is
\begin{equation}
n_s - 1  =  (2 - 4q_1)n  ~.
\end{equation}

\section{Summary and Conclusions} 
\label{sec6}

In this paper we have analyzed in detail the power spectrum of
curvature fluctuations in  a bouncing cosmology in which the bounce
phase has small vibrations, i.e., small bounces.  To be specific we
have mostly considered the case of one small bounce with
characteristic  time scales $\tau_B$ and $\Delta < \tau_B$ which are
much smaller than cosmological times.  We have given a detailed study
of the necessary matching conditions required to obtain the complete
form for the power spectrum. The matchings connect at least five
different phases for a given momentum scale which need to be treated
with care.

In our study, we have adopted two simplified models for the shape of
the vibrations, allowing a complete analytical study. Despite the
apparent simplicity of these models, they are already of sufficient
complexity to allow to extract similar features that can  emerge in
more realistic models. In particular,   similar structures that we
have considered here can appear in bounce models coming from  quantum
gravity, as those recently proposed in Ref.~\cite{Alesci}, which makes
this study of particular importance.  Our results for the power
spectrum shows that there is an amplification of its amplitude and it
also tends to get redder at large scales as the number of vibrations
increase. At small scales the power spectrum features superimposed
damped oscillations. 

The reddening of the spectrum for scales which enter the small bounce
agrees with the results found in Ref.~\cite{RHBcyclic}. The
oscillations in the power spectrum which are seen on small scales are
reminiscent of oscillations which are obtained in some other
approaches to the {\it Trans-Planckian problem} for cosmological
fluctuations. For example, if initial conditions are set on a
time-like {\it new physics hypersurface}~\cite{newphysics} such that
modes $k$ are initiated when the physical wavelength associated with
$k$ equals a fixed physical length (e.g. the Planck length), and they
are initiated in the same state (e.g.  the state which locally looks
like the Bunch-Davies vacuum~\cite{BD}), then oscillations in the
spectrum result.

Both the qualitative and quantitative changes in the power spectrum
that we have obtained  can produce observed effects in spectrum of
cosmological perturbations accessible through the measurements of the
cosmic microwave background radiation. These effects can manifest 
themselves both in pure bouncing cosmologies (no subsequent inflationary 
period) and in scenarios where there is a post-bounce
inflationary phase. {}For instance, those
bounce vibrations can induce particle production, changing the vacuum
state such as to be different from the usual Bunch-Davis one, similar
to recent pre-inflationary studies in Loop Quantum
Cosmology~\cite{Zhu:2017jew}. The modifications we have obtained in
this work could then be used to put constraints on these possible
features that can appear in  these bounce models and which deserve
further study. The results we have presented here provides then an
important first step in understanding these effects and which we hope
to address elsewhere.

\appendix

\section{Envelope of the Power Spectrum for Small Scale Modes}
\label{appA}

In this section, we would like to calculate the envelope of the power
spectrum for small scale modes. Since the model without plateau is a
special limit of the  model with a non-vanishing flat plateau, we just
focus on the latter.  We can simply set $\Delta \rightarrow 0$ to get
the answer for the model without a plateau.

By collecting the relevant terms in the power spectrum, we can write
it in the form
\begin{align}
P_{\zeta } = \left[C_1\sin(2k\Delta) + C_2 \cos(2k\Delta)\right]^2 P_i
\, .
\end{align}
The envelope of the power spectrum is thus
\begin{align}
P_{\zeta {\rm (env)}} = (C_1^2 + C_2^2 )P_i \, ,
\end{align}
where the coefficients $C_1$ and $C_2$ are given by
\begin{align}\nonumber
C_1&  = \frac{k^{-1}}{(1-2q_2)(1-2q_1)^2} \left\{ 2 \delta^{-1}
\left[k^2\delta^2+(q_1-1)q_1 \right](1+q_1-q_2)(-2+q_1+q_2)
\right. \\  & \left. - \delta^{2q_1-2} (q_1-k\delta) (q_1+k\delta)
(-2+q_1+q_2)^2 \epsilon^{1-2q_1} + \delta^{-2q_1}
(q_1-k\delta-1)(q_1+k\delta-1)(1+q_1-q_2)^2\epsilon^{2q_1-1} \right\}, 
\label{C1env}
\\\nonumber C_2 & =  \frac{1}{(1-2q_2)(1-2q_1)^2} \left[
  -2(1+q_1-q_2)(-2+q_1+q_2)  \right. \\ & \left. -\delta^{2q_1-1} 2
  q_1 (-2+q_1+q_2)^2 \epsilon^{1-2q_1} - \delta^{-2q_1+1} 2 (q_1-1)
  (1+q_1-q_2)^2 \epsilon^{2q_1-1} \right] \, .
\label{C2env}
\end{align}

We are interested in the parameter region $\epsilon / \delta \ll1$
(recall that the time scale $\epsilon$ is expected to be of the order
of the Planck scale, whereas $\delta$ will be parametrically larger
since it is associated with the time scale of the bounce). We are also
interested in the range of values $1/3<p<1$, or equivalently,
$1/2<q<+\infty$.  We can then determine which are the dominant terms
in $C_1$ and $C_2$, which from Eqs.~(\ref{C1env}) and (\ref{C1env}),
they are given by
\begin{align} 
C_1&  \simeq \frac{k^{-1}}{(1-2q_2)(1-2q_1)^2} \left[  -
  \delta^{2q_1-2} (q_1-k\delta) (q_1+k\delta)  (-2+q_1+q_2)^2
  \epsilon^{1-2q_1} \right], \\ C_2 & \simeq
\frac{1}{(1-2q_2)(1-2q_1)^2} \left[ -\delta^{2q_1-1} 2 q_1
  (-2+q_1+q_2)^2 \epsilon^{1-2q_1} \right] \, .
\end{align}

\begin{figure}[htbp]
\begin{center}
\subfigure[Model 1 with $\Delta=0.9 \tau_B$.]  {
  \includegraphics[width=0.43\textwidth]{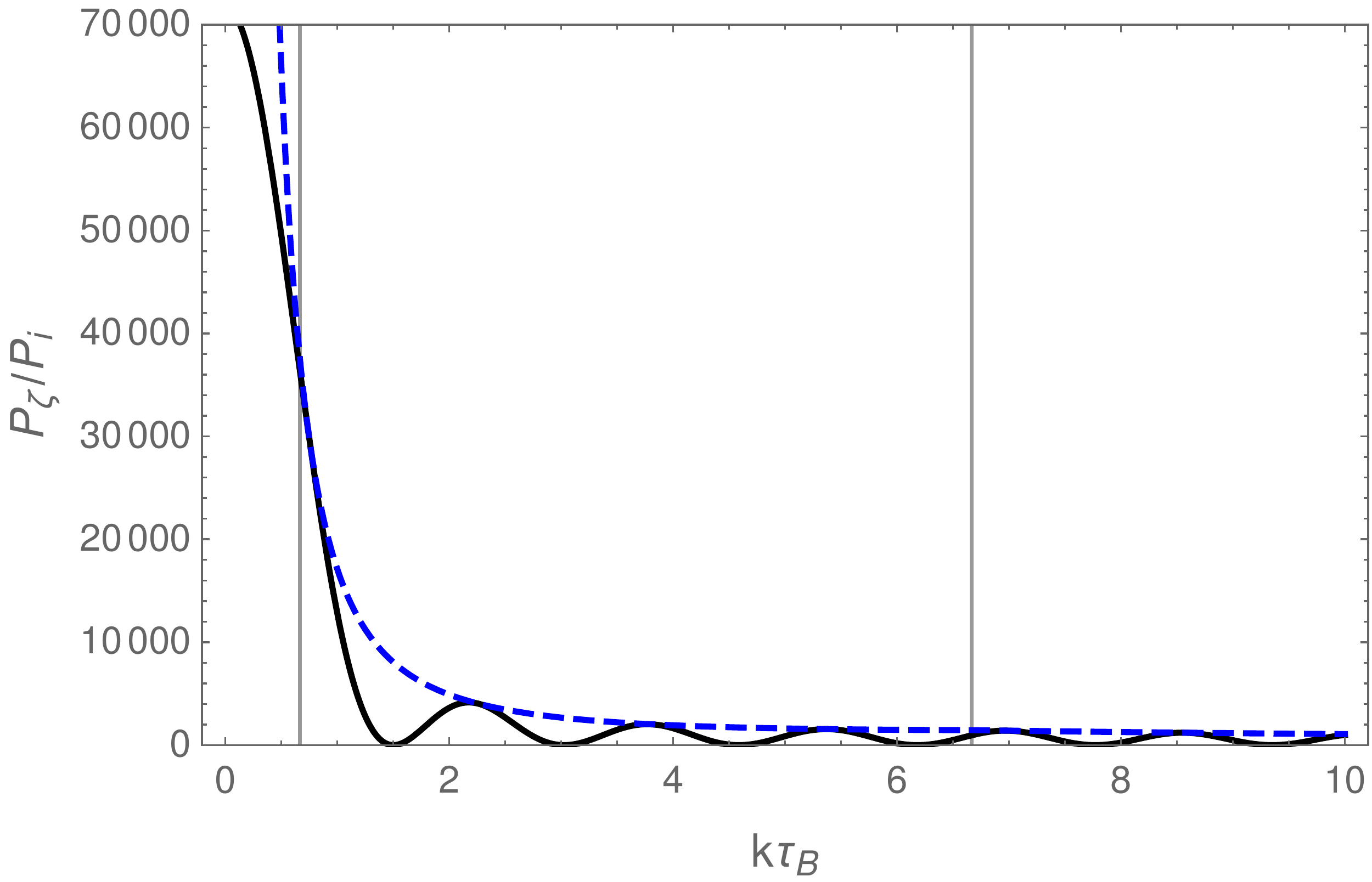}}
\subfigure[Model 2 ($\Delta=0$).]{
  \includegraphics[width=0.4\textwidth]{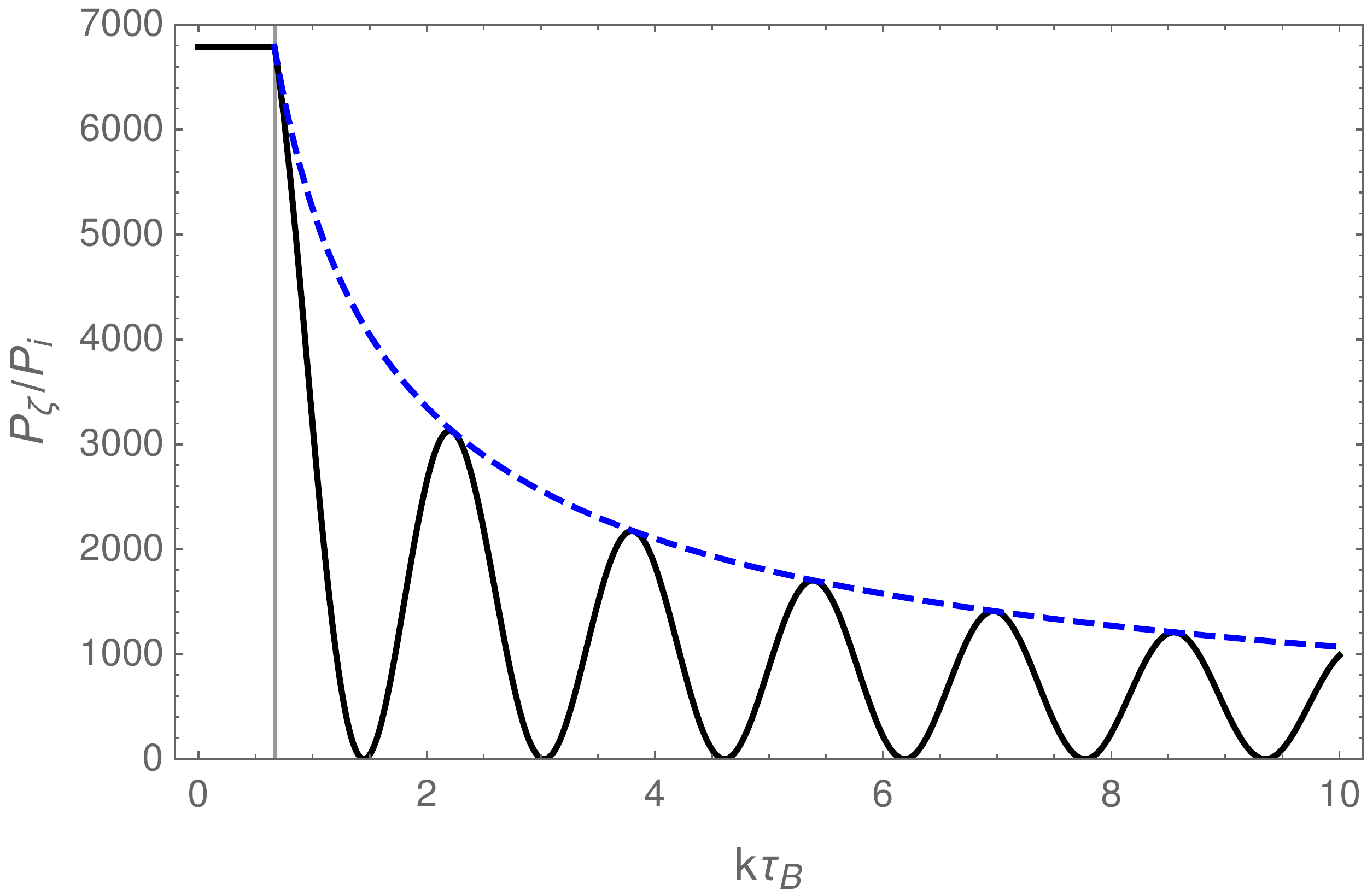}}
\end{center}
\caption{The normalized power spectrum and its envelop for the cases
  of the plateau model 1 (a)  and for the instantaneous model 2 (b),
  as a function of the comoving wavenumber times the bounce time
  $\tau_B$ and with parameters $q_1= q_2 = 2/3$ and $\epsilon = 0.01 \tau_B$.
  The two thin vertical lines in (a) indicates the two characteristic
  scales $k_{*,{\rm osc}}$ and $k_{*,{\rm flat}}$, while in (b) we
  only have the characteristic scale $k_{*,{\rm osc}}$.  }
\label{bounceplot3} 
\end{figure}

When $k$ is close to the $k_*$, then  $\delta \rightarrow
\tau_B-\Delta$ which is a constant.  In this case, taking the square
of $C_1$, we get terms with different with spectral indices, but the
dominant  contribution is the term with the lowest  power of $k$,
which the gives that the slope of the envelope (for an initial
spectrum which is scale-invariant) will be
\begin{align}
n_s -1 = -2 ,
\end{align}
because we have $k\delta<q_1$ in this range. This can be seen from the
numerical results shown in {}Fig.~\ref{bounceplot3} for the two models
we have considered. The change in the spectral slope is due to the
matching conditions. Each time, we can get factors of $1/k$ or $k$
when we match the solution across the boundaries of Regions II and
III, and of Regions III and IV. 

Note that in a generic case when we have a smooth evolution of the
scale factor,  we expect that there will be no discontinuities in the
power spectrum.  Thus, in a generic case, we do not expect that we
always get an interval of wavenumber with a spectrum of slope
$n_s=-2$. What we expect in the case of a smoothly evolving scale
factor is that on very large scales, we get a scale invariant
spectrum (the actual spectrum, not just the envelope), and then it
will smoothly transit  to a spectrum with tilt $n_s=-4q_1+2$ when we
look at the envelope only. We see  oscillations with amplitude given
by the envelope function  on intermediate and small scales.

The coefficient $C_2^2$ gives a scale invariant power spectrum
\begin{align}
n_s -1 = 0,
\end{align} 
but its amplitude is suppressed by $k\delta$ compared to the amplitude
of $C_1$.  To be a bit more precise (still in the case of constant
$\delta$), we can write
\begin{align}
C_1  = A_1 k^{-1} + A_2 k  \, ,
\end{align}
where the constants $A_1$ and $A_2$ are
\begin{align}\nonumber
A_1 & = \frac{1}{(1-2q_2)(1-2q_1)^2} \left[ 2 \delta^{-1} (q_1-1)q_1
  (1+q_1-q_2) (-2+q_1+q_2)  - \delta^{2q_1-2} q_1^2 (-2+q_1+q_2)^2
  \epsilon^{1-2q_1} \right. \\ &\left. -\delta^{-2q_1} (q_1-1)^2
  (1+q_1-q_2)^2 \epsilon^{2q_1-1}  \right],   \\ \nonumber A_2 & =
\frac{1}{(1-2q_2)(1-2q_1)^2} \left[ 2\delta(1+q_1-q_2)(-2+q_1+q_2)  +
  \delta^{2q_1} (-2+q_1+q_2)^2 \epsilon^{1-2q_1} \right.  \\ &- \left.
  \delta^{-2q_1+2} (1+q_1-q_2)^2 \epsilon^{2q_1-1}  \right] .
\end{align}
The spectral index is computed as
\begin{align}
n_s - 1 = \frac{d\ln P_{\zeta {\rm (env)}}}{d\ln k} = \frac{2
  P_i}{P_{\zeta {\rm (env)}}} (A_2^2 k^2 -A_1^2 k^{-2}).
\end{align}
The power spectrum is hence comprised of several terms with different
spectral tilts $n_s$
\begin{align}
n_s-1 = 2, 1, 0, -1, -2,
\end{align}
More generally (for larger values of $k$ when $\delta$ is not
constant), we have

\begin{align}
P_{\zeta } = \left\{D_1\sin[2(k\tau_B-q_1)] + D_2
\cos[2k(k\tau_B-q_1)]\right\}^2 P_i \, .
\end{align}
The envelope of the power spectrum is thus
\begin{align}
P_{\zeta {\rm (env)}} = (D_1^2 + D_2^2 )P_i \, ,
\end{align}
where the coefficients $D_1$ and $D_2$ are given by
\begin{align}
D_1 & = \frac{1}{(1-2q_2)(1-2q_1)} \left[ -2 (1+q_1-q_2) (-2+q_1+q_2)
  +  k^{2q_1-1} q_1^{-2q_1} (1+q_1-q_2)^2 \epsilon^{2q_1-1} \right] ,
\\ D_2 & = \frac{1}{(1-2q_2)(1-2q_1)^2} \left[
  -2(1+q_1-q_2)(-2+q_1+q_2) +  k^{1-2q_1} q_1^{2q_1-1}
  (-2q_1)(-2+q_1+q_2)^2\epsilon^{1-2q_1}  \right. \nonumber \\ &
  \left. -2 k^{2q_1-1} q_1^{-2q_1+1}  (q_1-1) (1+q_1-q_2)^2
  \epsilon^{2q_1-1} \right] \, .
\end{align}

We now can see that this envelope function reproduces the result of
Ref.~\cite{RHBcyclic}. We have
\begin{align}
D_1  &= B_1 + B_2 k^{2q_1-1},\\ D_2  &= E_1 + E_2 k^{1-2q_1} + E_3
k^{2q_1-1} \, ,
\end{align}
where the constants $B_1$, $B_2$, $E_1$, $E_2$ and $E_3$ are given by
\begin{align}
B_1  &= \frac{-2 (1+q_1-q_2) (-2+q_1+q_2)}{(1-2q_2)(1-2q_1)},\\ B_2
&= \frac{ q_1^{-2q_1} (1+q_1-q_2)^2 \epsilon^{2q_1-1}
}{(1-2q_2)(1-2q_1)},\\  E_1  &=
\frac{-2(1+q_1-q_2)(-2+q_1+q_2)}{(1-2q_2)(1-2q_1)^2},\\ E_2  &=
\frac{q_1^{2q_1-1}
  (-2q_1)(-2+q_1+q_2)^2\epsilon^{1-2q_1}}{(1-2q_2)(1-2q_1)^2}, \\ E_3
&= \frac{-2q_1^{-2q_1+1}  (q_1-1) (1+q_1-q_2)^2
  \epsilon^{2q_1-1}}{(1-2q_2)(1-2q_1)^2} .
\end{align}

The spectral tilt is then given by
\begin{align}
n_s - 1  & = \frac{P_i}{P_{\zeta {\rm (env)}}}  \left\{ 2 D_1 B_2
(2q_1-1) k^{2q_1-1} +  2 D_2 \left[E_2(1-2q_1) k^{-2q_1+1} + E_3
  (2q_1-1) k^{2q_1-1} \right]  \right\} \, .
\label{nsenv}
\end{align}
The expression (\ref{nsenv}) is comprised of several terms with
spectral tilts $n_s$ given by
\begin{align}
n_s-1 = 4q_1-2, 2q_1-1, 0, -4q_1+2, -2q_1+1 \, .
\end{align}
Since are interested in modes with $k\epsilon<1$ and parameter values
$1/3<p<1$ (or,  equivalently, $1/2<q<+\infty$) we can determine the
dominant terms in $D_1$ and $D_2$ and find them to be
\begin{align}
D_1  \rightarrow 0,\quad D_2  \rightarrow E_2 k^{1-2q_1} \, .
\end{align}
Thus, the dominant contribution to the power spectrum is
\begin{align}
P_{\zeta {\rm (env)} } = E_2^2 k^{2-4q_1} P_i \, ,
\end{align}
which corresponds to a spectral tilt of
\begin{align}
n_s = -4q_1+2 \, .
\end{align}

\section*{Acknowledgments}

One of us (R.B.) is grateful to Emanuele Alesci and Stefano Liberati
for discussions about the model of \cite{Alesci} which led to this
project. He also thanks Stefano Liberati and the other organizers of
the {\it Probing the Spacetime Fabric: from Concepts to Phenomenology}
workshop help in July 2017 at SISSA for inviting him to participate
and speak.
The research at McGill was supported in part by an NSERC Discovery
grant and by the Canada Research Chair program.  Q.L acknowledge
financial support from the University of Science and Technology of
China, and from the CAST Young Elite Scientists Sponsorship Program
(2016QNRC001), and by the NSFC (grant Nos. 11421303, 11653002).  SZ is
supported by the Hong Kong PhD Fellowship Scheme (HKPFS) issued by the
Research Grants Council (RGC) of Hong Kong.  R.O.R is partially
supported by research grants from Conselho Nacional de Desenvolvimento
Cient\'{\i}fico e Tecnol\'ogico (CNPq), grant No. 303377/2013-5 and
Funda\c{c}\~ao Carlos Chagas Filho de Amparo \'a Pesquisa do Estado do
Rio de Janeiro (FAPERJ), grant No.  E - 26/201.424/2014. Q.L.,
R.O.R. and S.Z. are grateful for the hospitality of the Physics
Department at McGill University during research visits when this work
was initiated.


\end{document}